\documentstyle[epsfig,twoside,fleqn,espcrc2]{article}
\newcommand{\msbar}{{\overline{\mbox{MS}}}}
\title{Next-to-leading order QCD corrections
to $e^+e^-\to 3$ jets \\ with massive quarks}
\author{A. Brandenburg\thanks{supported by Deutsche Forschungsgemeinschaft.}, W. Bernreuther, and P. Uwer
\\ \vskip 0.3cm Institut f\"ur Theoretische Physik,
RWTH Aachen, 52056 Aachen, Germany
\\ \vskip -6cm \hfill hep-ph/9709282   \vskip0.5cm
{\sf Talk given by A. Brandenburg at the QCD '97 conference, Montpellier, 3rd-9th July, 1997}}
\begin{document}
\begin{abstract}
\vskip 4.5cm
We discuss a calculation of the next-to-leading order QCD corrections
to the process $e^+e^-\to 3$ jets with massive quarks, and
show numerical results for the three jet fraction and the differential
two jet rate.  
\end{abstract}
%
\maketitle
\section{INTRODUCTION}
In view of the large number of jet events 
at the $Z$ resonance collected both at LEP
and SLC, it is desirable for precision tests of 
the standard model to use in the theoretical
predictions next-to-leading order (NLO) partonic matrix ele\-ments
that include the full quark mass dependence.
This is particularly important for $b$ quark enriched samples,
which can be obtained with high purity using vertex detectors. 
\par
The NLO QCD corrections to  $e^+e^-\to 3$ jets 
are well-known for massless quarks \cite{ElRoTe81,Fa81,Ve81,Ku81}. 
The three, four, and five jet rates involving massive quarks
have been computed to leading order in $\alpha_s$ already some
time ago \cite{Jo78,Al80}. Recently, results for the NLO
corrections to $e^+e^-\to 3$ jets
including mass effects have been reported
\cite{Ro96,BeBrUw97,BiRoSa97,NaOl97,BrUw97}.   
Knowing the NLO matrix elements including the mass dependence,
one can try to extract 
the mass of the $b$ quark from three jet rates involving $b$ quarks
at the $Z$ peak. This was
suggested in \cite{BiRoSa95}, elaborated in 
\cite{Ro96,BiRoSa97,Ro97}, and experimentally pursued
by the DELPHI collaboration \cite{Fu97,Fu97a}.
Further applications include precision tests of the 
asymptotic freedom property
of QCD by means of three jet rates and event shape variables
measured at various
center-of-mass energies, also far below the $Z$ resonance 
\cite{Be97}. For theoretical predictions 
concerning the production of top quark pairs at a future
$e^+e^-$ collider, the inclusion of the full mass dependence
is of course mandatory.
\par 
We have computed the complete differential
distributions for $e^+e^-$ annihilation into three and four partons
via a virtual photon or $Z$ boson at order $\alpha_s^2$,
including the full quark mass dependence \cite{BeBrUw97,BrUw97}. 
This allows for order $\alpha_s^2$
predictions of oriented three jet events (for the massless
case, see \cite{KoSc85}), and 
of any quantity that gets contributions only 
from three and four jet configurations.
In this talk we give an outline of our calculation (for 
details, see \cite{BrUw97}), after which we present some
numerical results.
\section{OUTLINE OF THE CALCULATION}
The calculation of an arbitrary quantity  
dominated by three jet configurations  
and involving a massive quark-antiquark pair 
to order $\alpha_s^2$ consists of two parts:
First, the computation of the amplitude of the partonic reaction
$e^+e^-\rightarrow\gamma^\ast, Z^\ast\rightarrow Q \bar Q  g$
at leading and next-to-leading order in the  QCD coupling.  
Here $Q$ denotes a massive quark
and $g$ a gluon. 
Second, the leading order matrix elements of the
four-parton production processes
$ e^+e^- \rightarrow Z^\ast, \gamma^\ast \rightarrow
ggQ\bar{Q},Q\bar{Q}q\bar{q},Q\bar{Q}Q\bar{Q}$ are needed.
Here $q$ denote light quarks which are taken to be massless.
\par
The infrared (IR) and ultraviolet (UV) singularities, which are encountered
in the computation of the one-loop integrals,
are treated within the framework
of dimensional regularization in $d=4-2\epsilon$ space-time dimensions.
We remove the UV singularities  
by the standard $\overline{\rm MS}$ renormalization.
We have converted from the outset the on-shell mass of the heavy quark $Q$
into the corresponding running $\overline{\rm MS}$ mass. It is known that far
from threshold one thereby absorbs some large logarithms into the running mass.
\par
After renormalization, the virtual corrections to 
the differential cross section for $e^+e^-\to Q\bar{Q}g$
still contain IR singularities. 
These have to be cancelled 
by the singularities that are obtained upon 
phase space integration of the squared tree amplitudes 
for the production of four partons.
Different methods to perform  this cancellation have been developed
(see \cite{GiGl92,FrKuSi96,CaSe96} and references therein).
We use the so-called phase space slicing method 
elaborated in \cite{GiGl92}, which we modified 
to account for masses \cite{BrUw97}. 
The basic idea is to ``slice'' the phase space of the
four parton final state by introducing an unphysical parton
resolution parameter $s_{min}$, which is much
smaller than all relevant physical scales. 
The parameter  $s_{min}$ splits the phase space into
a region where all four partons are ``resolved''
and a region where at least
one parton remains unresolved.
In the unresolved region soft and collinear
divergences reside, which have to be isolated explicitly to cancel
the singularities of the virtual corrections.
This is considerably simplified due to collinear and soft
factorizations of the matrix elements
which hold in the limit $s_{min}\to 0$.
(In the presence of massive quarks, the structure of
collinear and soft poles
is completely different as compared to the massless case.)
After having cancelled these IR poles  
against the IR poles of the
one-loop integrals entering the virtual corrections,
one is  left with a completely regular 
differential three-parton cross
section which depends on $s_{min}$.
The contribution to a three jet quantity of 
the ``resolved'' part of the four-parton cross section is finite
and may be evaluated in $d=4$ dimensions, which is
of great practical importance.
It also  depends on $s_{min}$ and is  most conveniently obtained 
by a numerical integration.
Since the parameter $s_{min}$ is
completely arbitrary, the sum of all contributions 
to any observable  must not depend on $s_{min}$.
Since the individual
contributions depend logarithmically
on $s_{min}$, it is a nontrivial test of the
calculation to demonstrate that
observables become independent of
$s_{min}$ for small values
of this parameter. 
\section{NUMERICAL RESULTS}
We will now discuss results for  
some three jet observables involving massive quarks.
All quantities are calculated by expanding in $\alpha_s$ 
to NLO accuracy.
We consider here the JADE \cite{Ba86} and Durham \cite{Br90}
jet finding algorithms, 
although other schemes \cite{BeKu92} can also be
easily implemented. 
First we demonstrate the independence
of physical quantities on the parameter $s_{min}$ as $s_{min}\to 0$. 
\begin{figure}[t]
\begin{center}
\epsfig{file=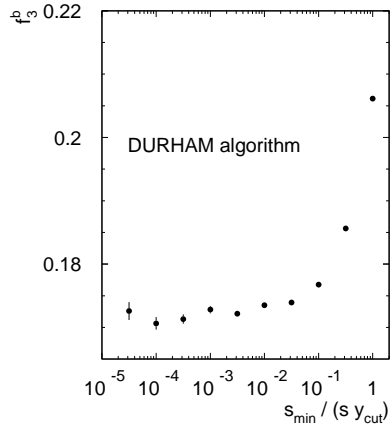,height=6cm}\vskip-1.2cm 
\caption{\small
The three jet fraction 
$f_3^b$ at NLO as defined in the text at 
$\protect\sqrt{s}=\mu=m_Z$ as a function of 
$y_{min}=s_{min}/(sy_{cut})$
for the Durham algorithm at a value of the jet resolution
parameter $y_{cut}=0.03$ with 
$m^{\mbox{\scriptsize $\msbar$}}_b(\mu=m_Z)=3$ GeV and 
$\alpha_s(\mu=m_Z)=0.118$.
\vskip -1cm}
\label{fig:smindu}
\end{center}
 \end{figure}
We choose
as an example the three jet fraction for $b$ quarks,
\begin{eqnarray} 
\label{f3b}
f_3^b(y_{cut})=\frac{\sigma_3^b(y_{cut})}{\sigma_{tot}^b}.
\end{eqnarray}
In (\ref{f3b}), the numerator $\sigma_3^b$ 
is defined as the three jet cross section
for events in which at least two jets containing a $b$ or $\bar{b}$ quark
remain after the clustering procedure. This requirement  ensures that
the cross section stays finite also in the limit $m_b\to 0$. 
The contribution of the process
$e^+e^-\to Z,\gamma^{*}\to q\bar{q}g^{*}\to q\bar{q}b\bar{b}$ to the three
jet cross section with {\it one} tagged $b$ quark develops 
large logarithms $\ln(m_b^2)$ 
-- which find no counterpart in the virtual corrections
against which they can cancel --
when the $b\bar{b}$ pair is clustered into a single jet. 
In principle, there are three ways to handle this problem: 
One may  impose suitable experimental requirements/cuts to suppress
contributions from events with  
two light quark jets and one jet containing a $b\bar{b}$
pair (the definition for $\sigma_3^b$
chosen by us is an example for this), or one can 
improve the fixed order calculation by absorbing the large logarithm
into a fragmentation function for a gluon into a $b$ quark.
The third possibility is simply to keep the large $\ln(m_b^2)$ term.
This may however lead to an overestimate of the $b$ quark
mass effects in some observables.
A detailed discussion of this issue will be presented elsewhere 
\cite{BeBrUw297}. 
\begin{figure}[t]
 \begin{center}
\epsfig{file=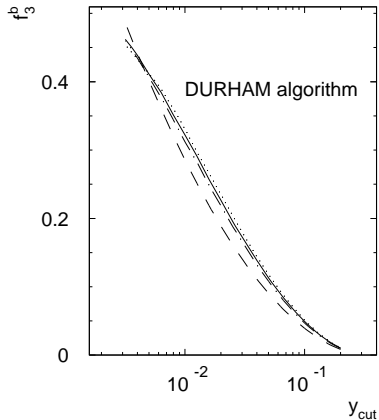,height=6cm}\vskip-1.2cm
\caption{\small
The three jet fraction 
$f_3^b$ as defined in the text at 
$\protect\sqrt{s}=m_Z$
as a function of $y_{cut}$ for the Durham algorithm. 
The dashed line is the LO result.
The NLO results are for $\mu=m_Z$ (solid line),
$\mu=m_Z/2$ (dotted line), and $\mu=2m_Z$ (dash-dotted
line). 
\vskip -1cm}
\label{fig:ycutdu}
\end{center}
 \end{figure}
Fig.  \ref{fig:smindu} shows the three jet 
fraction $f_3^b$ in the 
Durham  scheme at NLO as a function
of $y_{min}=s_{min}/(sy_{cut})$ at a fixed  value of $y_{cut}=0.03$ 
and $\sqrt{s}=m_Z=91.187$ GeV. 
The corresponding plot for the JADE scheme
has a  similar shape.
For the renormalization scale we take in this
plot $\mu=\sqrt{s}$. 
As mentioned above, we use 
$m^{\mbox{\scriptsize $\msbar$}}_b(\mu)$ defined
in the $\msbar$ scheme at the scale $\mu$. 
The asymptotic
freedom property of QCD predicts that this mass 
parameter decreases when being evaluated at a higher
scale.  
(For low energy determinations of the $b$ quark mass
 see e.g. \cite{Ne94,Na94} and references therein.)
With $m^{\mbox{\scriptsize $\msbar$}}_b(\mu=m_b)=4.36$ GeV 
\cite{Ne94}
and $\alpha_s(m_Z)=0.118$ as an input and employing 
the standard renormalization group evolution of the 
coupling and the quark masses, 
we use the value $m^{\mbox{\scriptsize $\msbar$}}_b(\mu=m_Z)= 3$ GeV.
One clearly sees that $f_3^b$ reaches a plateau
for small values of $y_{min}$. The error in the numerical integration
becomes bigger as $y_{min}\to 0$.  In order to keep this error as small as 
possible without introducing a systematic error from using the soft and collinear
approximations, we take in the following 
$y_{min}=0.5\times 10^{-2}$ for the Durham algorithm and
$y_{min}=10^{-2}$ for the JADE algorithm. 
At these values, 
the dominant $s_{min}$-dependent individual contributions from three and 
four resolved partons are about
a factor of 4 (Durham) and  2.5 (JADE) larger than the sum. 
In Fig.  \ref{fig:ycutdu} 
we plot  $f_3^b$ as a function of 
$y_{cut}$, again at $\sqrt{s}=m_Z$.
The QCD corrections to the LO result are quite 
sizable as known also in the massless case.
The renormalization scale dependence (where $\mu$ is varied
between $m_Z/2$ and $2m_Z$), which is also shown in
Fig. \ref{fig:ycutdu}, is modest in the whole $y_{cut}$ 
range exhibited for the Durham algorithm. 
\begin{figure}[t]
 \begin{center}
\epsfig{file=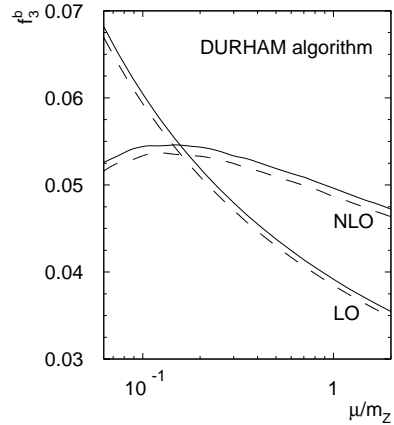,height=6cm}\vskip-1.2cm
\caption{\small Dependence of the three jet fraction $f_3^b$ for
the Durham algorithm
on the renormalization scale at $\protect\sqrt{s}=m_Z$ 
and $y_{cut}=0.2\times 10^{-3/10}\approx 0.1$
for on-shell $b$ quark masses $m_b^{\rm pole}=3$ GeV 
(full curves) and
$m_b^{\rm pole}=5$ GeV (dashed curves).
\vskip -1cm}
\label{fig:renormdu}
\end{center}
\end{figure}
In Fig. \ref{fig:renormdu} we take a closer look
on the scale dependence of $f_3^b$, now using the on-shell mass 
renormalization scheme. We vary the scale $\mu$ between $m_Z/16$ and
$2m_Z$ for a fixed value $y_{cut}=0.2\times 10^{-3/10}\approx 0.1$ and on-shell masses
$m^{\rm pole}_b=3$ GeV and $m^{\rm pole}_b=5$ GeV. 
We see that the scale dependence of
the LO result (which is solely due to
the scale dependence of $\alpha_s$ at this order) 
in the Durham algorithm  amounts
to about 100\% in the $\mu$ interval shown. The 
inclusion of the $\alpha_s^2$ corrections reduces the scale dependence
significantly; the NLO result for $f_3^b$ at $\mu=2m_Z$ is about 10\% smaller
than the NLO result at $\mu=m_Z/16$. 
In the case of the JADE algorithm, which we do not show here, 
the difference between 
$f_3^b$ at $\mu=2m_Z$ and at $\mu=m_Z/16$ is reduced from
about 100\% at LO to about 30\% at NLO.         
\par
The effect of the $b$ quark mass may be illustrated by
looking at the double ratio 
\begin{eqnarray}
{\cal C}(y_{cut})=\frac{f_3^b(y_{cut})}{f_3^{{\rm incl.}}(y_{cut})},
\end{eqnarray}
where the denominator is the three jet fraction when summing
over all active quark flavors, which is given to a very good 
approximation by the massless NLO result \cite{KuNa89}. 
Similar double ratios have been studied
in \cite{Ro96,BeBrUw97,BiRoSa97}, and \cite{BiRoSa95,Ro97}.
\begin{figure}[t]
 \begin{center}
\epsfig{file=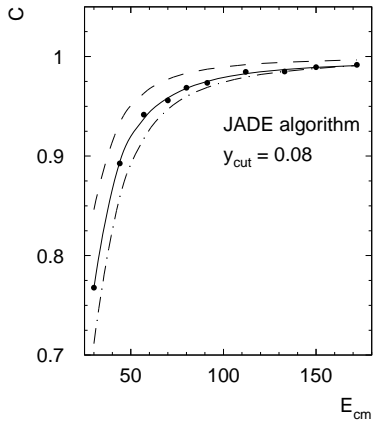,height=6cm}\vskip-1.2cm
\caption{\small The double ratio ${\cal C}$ 
as a function of the c.m. energy 
at $y_{cut}=0.08$ for the JADE algorithm.
The full curve and the points show the NLO result, the dashed curve shows the LO 
result. In both cases a running
mass $m^{\mbox{\scriptsize $\msbar$}}_b(\mu=\protect\sqrt{s})$
evolved from $m^{\mbox{\scriptsize $\msbar$}}_b(\mu=m_Z)=3$ GeV is used.
The dash-dotted curve shows the LO result for a fixed  mass 
$m_b=4.7$ GeV.\vskip -1cm}
\label{fig:edep}
\end{center}
 \end{figure}
In Fig. 
\ref{fig:edep}  we
plot ${\cal C}$ as a function of the c.m. energy at $y_{cut}=0.08$
for the JADE algorithm. The running of $\alpha_s$  
is taken into account in the
curves, where we again use as an input $\alpha_s(\mu=m_Z)=0.118$.
For the dashed (LO) and full (NLO) curve we use a running 
mass  $m^{\mbox{\scriptsize $\msbar$}}_b(\mu=\sqrt{s})$ evolved
from $m^{\mbox{\scriptsize $\msbar$}}_b(\mu=m_Z)=3$ GeV. 
For comparison we
also show the LO result for a fixed value of the $b$ quark mass 
$m_b=4.7$ GeV (dash-dotted curve), which is the corresponding
value of the pole mass. One clearly sees that the effect 
of the $b$ quark mass gets larger for smaller c.m. energies.
\par
\begin{figure}[t]
 \begin{center}
\epsfig{file=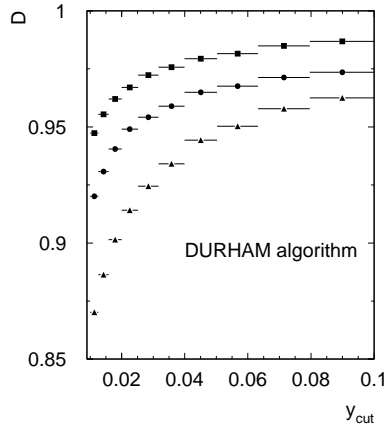,height=6cm}\vskip-1.2cm
\caption{\small The double ratio ${\cal D}$ as a function
of $y_{cut}$
for $\protect\sqrt{s}=\mu=m_Z$.
Full circles: NLO results for 
$m^{\mbox{\scriptsize $\msbar$}}_b(\mu=m_Z)=3$ GeV. 
For comparison, the
squares (triangles) are the LO results for
$m_b=3$ GeV ($m_b=5$ GeV). The horizontal bars
show the size of the bins in $y_{cut}$.
\vskip-1cm}
\label{fig:d2du}
\end{center}
\end{figure}
Another interesting quantity to study mass effects is the 
differential two jet rate \cite{Op90} defined as
\begin{equation}
D_2(y)=\frac{f_2(y)-f_2(y-\Delta y)}{\Delta y},
\end{equation}
where $f_2(y)$ is the two jet fraction at $y=y_{cut}$ for
a given jet algorithm. The advantage of $D_2$ over the
three jet fraction $f_3$ lies in the fact that the statistical
errors in bins of $D_2(y)$ are independent from each other
since each event enters the distribution only once. To order
$\alpha_s^2$, $D_2(y)$ can be calculated from the three- and
four jet fractions using the identity 
\begin{equation}
1=f_2+f_3+f_4+O(\alpha_s^3).
\end{equation}
We define
\begin{equation}
{\cal D}(y)=\frac{D_2^b(y)}{D_2^{\rm incl}(y)},
\end{equation}
where we -- as in the case of the quantity ${\cal C}$ -- 
use the massless NLO result to evaluate
the denominator. 
We plot our result for ${\cal D}(y)$
in the Durham scheme in Fig. \ref{fig:d2du}, again
for $\sqrt{s}=\mu=m_Z$. 
The effects of the $b$ quark mass are of the order
of 5\% or larger at small values of $y_{cut}$.
\section{CONCLUSIONS}
%
We have presented NLO 
results for a number of three jet observables 
involving massive quarks. It will be interesting to see how 
our predictions compare to detailed experimental
analyses.

\section*{DISCUSSIONS}
\noindent {\bf A.P. Contogouris}\newline
{\it You stated three ways for treating large logarithms and that you followed
one of them. Suppose you follow any of the other two. How much your 
higher corrections will change? Do you have any idea about their
sensitivity, perhaps from another process?}\newline
\ \newline
{\bf A. Brandenburg}\newline
{\it If one adds the contribution from $g^{\ast}\to b\bar{b}$ splitting ''naively'', i.e.
keeps the large logarithm in the three jet cross section with one tagged $b$ 
quark as it is, the ratio ${\cal C}(y_{cut})$ evaluated at 
$\sqrt{s}=\mu=m_Z$ is, for large values of $y_{cut}$,  up to
3\% larger as compared to the result 
for this ratio with two tagged $b$ quark jets. 
The difference decreases for smaller values of 
$y_{cut}$.}

\end{document}